\documentclass[reprint,aps,pre,amsmath,amssymb,superscriptaddress]{revtex4-2}
\usepackage{graphicx}
\usepackage{dcolumn}
\usepackage{bm}
\usepackage{xcolor}
\usepackage{float}
\usepackage[normalem]{ulem}

\begin{document}

\preprint{APS/123-QED}

\title{Activity-induced droplet propulsion and multifractality}

\author{Nadia Bihari Padhan}
 \email[]{nadia@iisc.ac.in}
\author{Rahul Pandit}
 \email[]{rahul@iisc.ac.in}
\affiliation{
 Centre for Condensed Matter Theory, Department of Physics, Indian Institute of Science, Bangalore 560012, India 
}

\date{\today}

\begin{abstract}
 We develop a minimal hydrodynamic model, without an orientational
 order parameter, for assemblies of contractile swimmers encapsulated in a droplet of a binary-fluid emulsion. Our model uses two coupled scalar order parameters, $\phi$ and $\psi$, which capture, respectively, the droplet interface and the activity of the contractile swimmers inside this droplet. These order parameters are also coupled to the velocity field $\bm u$. At low activity, our model yields a self-propelling droplet whose center of mass $(CM)$ displays rectilinear motion, powered by the spatiotemporal evolution of the field $\psi$, which leads to a time-dependent vortex dipole at one end of the droplet. As we increase the activity, this $CM$ shows chaotic super-diffusive motion, which we characterize by its mean-square displacement; and the droplet interface exhibits multifractal fluctuations, whose spectrum of exponents we calculate. We explore the implications of our results for experiments on active droplets of contractile swimmers.
\end{abstract}

\maketitle

\section{Introduction}
Active matter comprises systems that are far from equilibrium and in which the constituents extract energy from their surroundings, do mechanical work, and dissipate energy to the same environment~\cite{ramaswamy2017active,bowick2022symmetry,marchetti2013hydrodynamics,mahault2018outstanding}. The self-organisation of the constituents of such systems can lead to large-scale pattern formation, observed in, e.g., crowds~\cite{castellano2009statistical,bottinelli2016emergent}, fish schools~\cite{becco2006experimental},
bird flocks~\cite{bialek2012statistical,cavagna2010scale}, and bacterial colonies~\cite{chen2017weak}. They exhibit a variety of fascinating emergent phenomena, e.g., Motility-Induced-Phase-Separation (MIPS), in which an initially uniform state of active swimmers separates spontaneously into dense and dilute phases, driven by persistent motion and repulsion ~\cite{wittkowski2014scalar,cates2015motility,gonnella2015motility}.
Most experiments, with motile bacteria or synthetic micro-swimmers, use confinement -- solid immovable or soft, e.g., by
a droplet boundary --  that plays a crucial role in the spatiotemporal patterns in assemblies of active micro-swimmers~\cite{wioland2013confinement, huang2021emergent, ramos2020bacteria}.
Certain bacterial systems, when confined to a droplet, can propel and deform the droplet~\cite{kokot2022spontaneous, sokolov2018instability, ramos2020bacteria}; self-propelling, or \textit{active} droplets
have been considered in nematically ordered, active-polar, chemically driven, and phase-field systems~\cite{gao2017self,vcopar2019topology,tjhung2012spontaneous,de2014spontaneous,a2014active, yoshinaga2019self, ruske2021morphology,fadda2017lattice,singh2020self}.

We develop a \textit{minimal phase-field model} for assemblies of \textit{contractile swimmers} encapsulated in a droplet of a binary-fluid emulsion, to obtain self-propelling droplets, which are powered by the rich spatio-temporal dynamics of the contractile-swimmer field; this propulsion does not require any orientational ordering. Our results are of direct relevance to active droplets of contractile swimmers such as \textit{Chlamydomonas reinhardtii}~\cite{yeomans2014introduction,fragkopoulos2021self} (\textit{C. reinhardtii}). 
Our phase-field theory has two conserved scalar order parameters $\phi$ and $\psi$. The former distinguishes between two, coexisting liquid phases, separated by an interface at the droplet boundary; the latter is related to the bacterial concentration; $\phi$ and $\psi$ are coupled to each other and to the flow velocity $\bm u$, as in the Cahn-Hilliard-Navier-Stokes (CHNS) system or model H~\cite{pal2016binary,perlekar2017two}. In the absence of the direct coupling between the two order parameters, our model decouples into (a) the CHNS model, for $\phi$ and $\bm u$, that has been used to study mutifractal droplet dynamics~\cite{pal2016binary} in a turbulent flow and (b) the \textit{active model} H, for $\psi$ and $\bm u$, that has been used to study MIPS~\cite{wittkowski2014scalar,Tiribocchi2015,shaebani2020computational}.

We carry out pseudospectral direct numerical simulations (DNSs) of our model to uncover the dependence of the spatiotemporal evolution of an emergent active droplet on the \textit{activity} parameter $A$ (defined below). For low values of $A$, the center-of-mass $(CM)$ of the droplet shows rectilinear motion, associated with a time-dependent vortex dipole at one of its ends; as $A$ increases, the droplet
fluctuates and its $CM$ exhibits a crossover from rectilinear to super-diffusive motion, reminiscent of L\'evy walks. Furthermore, at large values of $A$, the bacterial field generates low-Reynolds-number, but turbulent, flows and \textit{multifractal deformation} of the active-droplet boundary. 

\section{Model}
We use the free-energy functional
\begin{widetext}
\begin{eqnarray}
\mathcal F[\phi, \nabla \phi, \psi, \nabla \psi] = \int_{\Omega} \left(\frac{3}{16}\left(\frac{\sigma_1}{\epsilon_1}(\phi^2-1)^2 + \frac{\sigma_2}{\epsilon_2} (\psi^2-1)^2 \right) - \beta \phi \psi  +
\frac{3}{4}\left( \sigma_1 \epsilon_1 |\nabla \phi|^2 + \sigma_2 \epsilon_2 |\nabla \psi|^2\right) \right) d\Omega\,,
\label{eq:fren}
\end{eqnarray}
\end{widetext}
where $\Omega$ is the region we consider, $\sigma_1$ and $\sigma_2$ are surface-tension coefficients, $\epsilon_1$ and $\epsilon_2$ are widths 
of the $\phi$ and $\psi$ interfaces, respectively, and the attractive coupling $\beta > 0$. To address experiments on active droplets carried out under planar confinement we use the following 2D \textit{active} CHNS equations:
\begin{eqnarray}
\partial_t \phi + (\bm u \cdot \nabla) \phi &=& M_1 \nabla^2 \left( \frac{\delta \mathcal F}{\delta \phi}\right)\,; \label{eq:phi}\\
\partial_t \psi + (\bm u \cdot \nabla) \psi &=& M_2 \nabla^2 \left( \frac{\delta \mathcal F}{\delta \psi}\right)\,;\label{eq:psi}\\
\partial_t \omega + (\bm u \cdot \nabla) \omega &=& \nu \nabla^2 \omega -\alpha \omega +[\nabla \times (\mathfrak{S}^{\phi} +  \mathfrak{S}^{\psi})]\,;\label{eq:omega}\\
\nabla \cdot \bm u &=& 0\,; \quad \omega = (\nabla \times \bm u)\,;\label{eq:incom}\\
\mathfrak{S}^{\phi} &=& -(3/2)\sigma_1 \epsilon_1 \nabla^2 \phi \nabla \phi \,;\label{eq:Sphi}\\
 \mathfrak{S}^{\psi} &=&    -(3/2) \tilde\sigma_2 \epsilon_2 \nabla^2 \psi \nabla \psi\,; \label{eq:Spsi}
\end{eqnarray} 
 the constant fluid density $\rho=1$, the advection-diffusion equations~(\ref{eq:phi}) and (\ref{eq:psi}) use the constant mobilities $M_1$ and $M_2$ for $\phi$ and $\psi$, respectively, and the 2D incompressible Navier-Stokes equations~(\ref{eq:omega}) and (\ref{eq:incom}) use the vorticity $\omega$, the kinematic viscosity $\nu$, and the bottom friction $\alpha$; the interfacial stress $\mathfrak{S}^{\phi}$ [Eq.~(\ref{eq:Sphi})] from $\phi$ and is derived from $\mathcal F$; for the \textit{active stress} $\mathfrak{S}^{\psi}$ [Eq.~(\ref{eq:Spsi})] from $\psi$ we use the active-model-H formulation for MIPS~\cite{wittkowski2014scalar,Tiribocchi2015,shaebani2020computational}; both $\omega$ and 
$[\nabla \times (\mathfrak{S}^{\phi} +  \mathfrak{S}^{\psi})]$ lie normal to the 2D plane.
We refer to $\psi$ as the active scalar ~\footnote{The active-matter terminology and the conventional fluid-dynamics nomenclature are slightly different. In the fluid-dynamics sense, both $\phi$ and $\psi$ are active scalars insofar as they affect the velocity field $\bm u$. However, in the active-matter sense, $\psi$ is active but $\phi$ is not.}. Note that the mechanical surface tension $\tilde \sigma_2 \neq \sigma_2$;
and $\tilde \sigma_2$ can take both negative and positive values unlike $\sigma_1$ and $\sigma_2$, which are always positive. For contractile (extensile) swimmers $\tilde \sigma_2 < 0 \,\,(\tilde \sigma_2 > 0)$ and the system shows arrested phase separation (complete phase separation) ~\cite{Tiribocchi2015}. The spatiotemporal evolution of the fields in Eqs.~(\ref{eq:fren})-(\ref{eq:Spsi}) depend on the initial conditions (see below) and the non-dimensional Cahn numbers $\text{Cn}_1 = \epsilon_1/R_0$ and $\text{Cn}_2 = \epsilon_2 / R_0$, Weber numbers $\text{We}_1 = R_0 U_0^2 /\sigma_1$ and $\text{We}_2 = R_0 U_0^2/\sigma_2$, Peclet numbers $\text{Pe}_1 = R_0 U_0 \epsilon_1/(M_1 \sigma_1)$ and $\text{Pe}_2 = R_0 U_0 \epsilon_2/(M_2 \sigma_2)$, order-parameter couplings $\beta_1^{\prime} = \beta \epsilon_1/\sigma_1$ and $\beta_2^{\prime} = \beta \epsilon_2 / \sigma_2$, friction $\alpha^{\prime} = \alpha R_0 / U_0$, Reynolds number $\text{Re} = R_0U_0/\nu$, where $U_0 = {\left< U_{CM}(t) \right>_t}$, with $U_{CM}$
the speed of the droplet's center of mass ($CM$) (see below and Appendix ~\ref{app:non-dimensions}), and, most important, the \textit{activity}  
\begin{equation}
A= |\tilde \sigma_2|/\sigma_2\,;
\end{equation}
 we concentrate on contractile swimmers with $\tilde \sigma_2 < 0$. In Table-I of Appendix ~\ref{app:non-dimensions} we list the parameters for our DNS runs R1-R7.
 
 \section{Initial conditions and numerical methods}
 We consider an initially stationary and circular droplet, of radius $R_0$ and with its center at $(x_{0,1}, x_{0,2}) = (\pi, \pi)$:
\begin{eqnarray}
\bm{u}(\bm{x},t=0)&=&0\,;\nonumber \\
\phi(\bm x, t=0) &=& \tanh{\left(\frac{R_0 - \sqrt{(x_1-x_{0,1})^2 + (x_2-x_{0,2})^2}}{\epsilon_1}\right)}\,;\nonumber \\
    \psi(\bm x, t=0) &=& \begin{cases}
    \psi_0(\bm x) &\text{for}\quad |\bm x|\leq R_0 \,;\\
    -1 &\text{for}\quad |\bm x| > R_0 \,;
    \label{eq:init}
    \end{cases}
\end{eqnarray}
 $\psi_0 (\bm x)$ is a random number distributed uniformly on the interval $[-0.1,0.1]$. Regions with negative (positive) values of $\phi$ and $\psi$ have low (high) densities of these scalars. 

 Our DNS of Eqs.~(\ref{eq:fren})-(\ref{eq:Spsi}) employs a standard Fourier pseudospectral method~\cite{canuto2012spectral}, with the $1/2$ rule for the removal of aliasing errors. We use a square domain of side $L = 2\pi$, with period boundary conditions in both spatial directions, and $N^2$ collocation points. For time integration, we use the semi-implicit ETDRK-2 method~\cite{cox2002exponential}. Our computer program is written in CUDA C and is optimised for recent GPU architectures, such as the one used in the NVIDIA A100 processor.

\section{Spatiotemporal evolution}
\begin{figure}
{
\includegraphics[width=8.5cm]{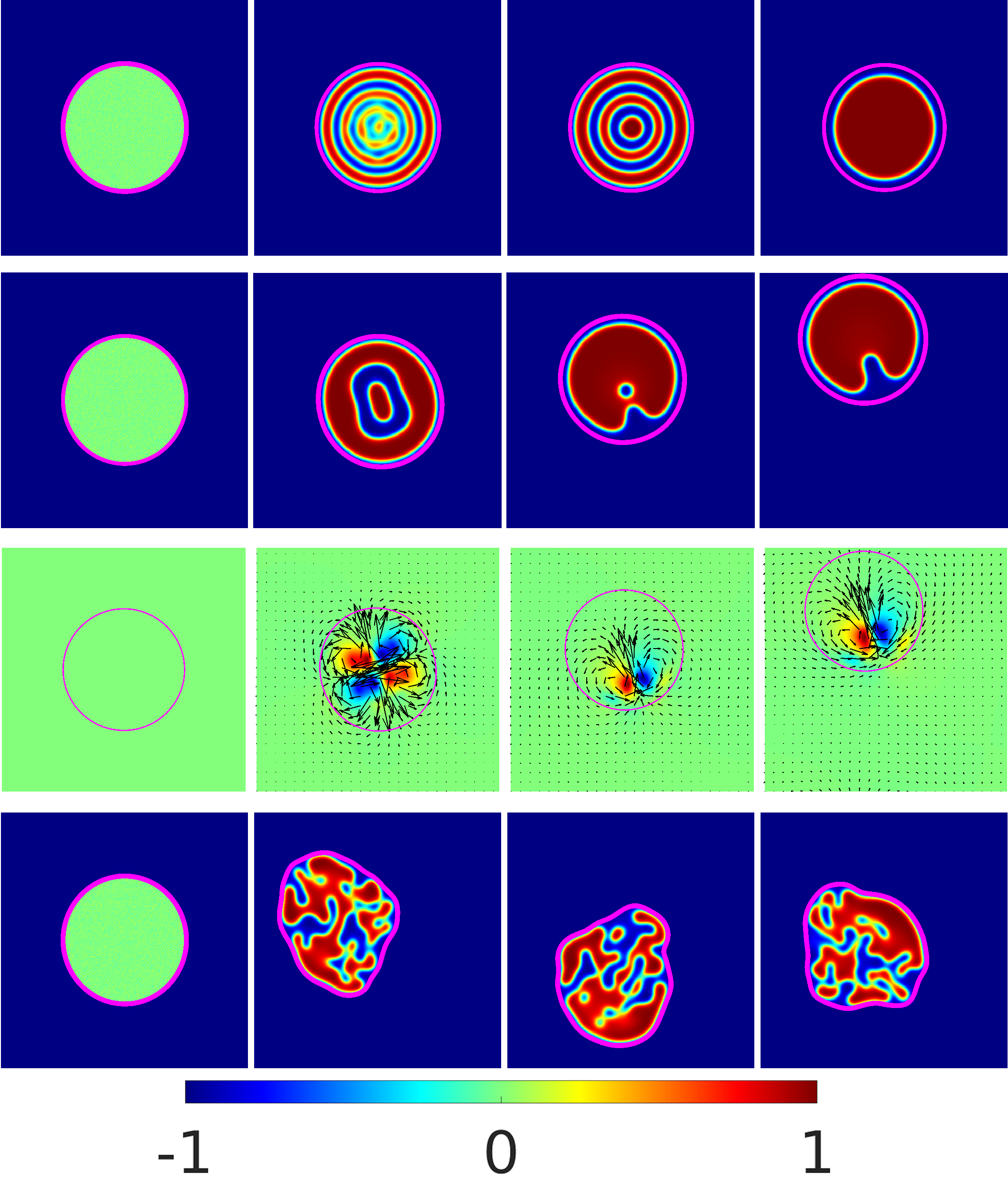}
\put(-200,300){\rm {$\xrightarrow[\textit{\bf{time}}]{\hspace*{5cm}}$}}
\put(-260,250){\rm {\bf(a)}}
\put(-260,190){\rm {\bf(b)}}
\put(-260,123){\rm {\bf(c)}}
\put(-260,60){\rm {\bf(d)}}
}

\caption{\label{fig:phi_psi} Illustrative pseudocolor plots of $\psi$, with the $\phi = 0$ contour shown in magenta, at different representative times (increasing from left to right) for (a) $A = 0$ (no droplet propulsion), (b) $A = 0.15$ (rectilinear droplet propulsion), and (d) $A = 1$ (turbulent droplet propulsion). In (c) we show, for $A = 0.15$, vector plots of the velocity field $\bm u$, with the $\phi = 0$ contour line (magenta), overlaid on a pseudocolor plot of the vorticity $\omega$ normalised by its maximal value; the lengths of velocity vectors are proportional to their magnitudes. [See videos V1, V2, V3, and V4, in Appendix ~\ref{app:videos}.]}
\end{figure}

 To monitor the spatiotemporal evolution of the initial droplet [Eq.~(\ref{eq:init})] we obtain pseudocolor plots of $\phi$ and $\psi$ (Fig.~\ref{fig:phi_psi}) and  we compute
\begin{eqnarray}
U_{CM}(t)&=& \sqrt{\sum^2_{i=1}\left[\sum_{\bm x \ni {\phi(\bm x, t) > 0}} u_i(\bm x,t)\right]^2}\,,\nonumber \\
E(k,t)&=& \frac{1}{2}\displaystyle \sum_{k-1/2<k'<k+1/2} \left[\hat{ \bm{u}}(\mathbf{k}',t)\cdot \hat{ \bm{u}}(-\mathbf{k}',t)\right]\,,\nonumber \\
\text{S}_\psi(k,t) &=& \displaystyle \sum_{k-1/2<k'<k+1/2} |\hat \psi(\mathbf{k}', t)|^2\,, \nonumber\\
\text{S}_\phi(k,t) &=& \displaystyle \sum_{k-1/2<k'<k+1/2} |\hat \phi(\mathbf{k}', t)|^2\,, \nonumber\\
\mathcal L (t) &=& 2\pi \sum_{k} \text{S}_\psi(k,t)/\sum_{k} k \text{S}_\psi(k,t)\,,\nonumber\\
\mathcal{M}(t) &=& \left<\displaystyle \sum_{i=1}^2 (x_{CM,i}(t) - x_{CM,i}(t_0))^2\right>\,, \nonumber\\
\Gamma(t) &=& \left[\mathcal{S}(t)/\mathcal{S}_0(t)\right] - 1\,,
\label{eq:spatem}
\end{eqnarray}
which are, respectively, the speed of the droplet's $CM$, the fluid energy spectrum, the spectra of $\psi$ 
and $\phi$, a length scale that follows from $\text{S}_\psi$, the mean-square displacement of the droplet's $CM$, and the normalised perimeter of the $\phi=0$ contour that bounds the droplet [$\mathcal{S}(t)$ is the perimeter of the droplet at time $t$ and $\mathcal{S}_0(t)$ is the perimeter of a circular droplet of equal area at time $t$]; the subscripts $i$ and $CM$ denote Cartesian components and the droplet's $CM$, respectively; carets indicate spatial Fourier transform; and $k$ and $k'$ are the moduli of the wave vectors $\mathbf{k}$ and $\mathbf{k}'$.
\begin{figure*}
{\includegraphics[width=18.5cm]{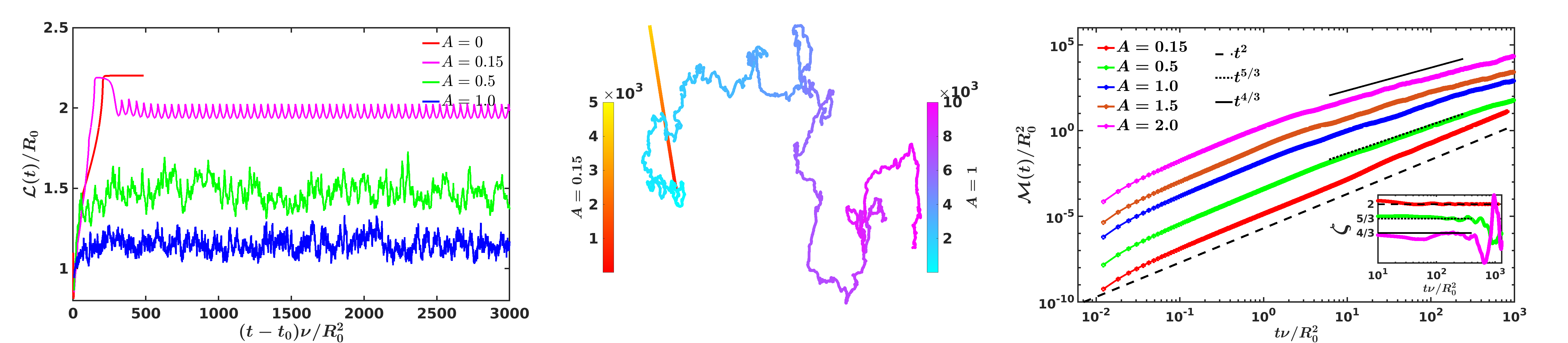}
\put(-525,100){\rm {\bf(a)}}
\put(-330,100){\rm {\bf(b)}}
\put(-190,100){\rm {\bf(c)}}
}

\caption{\label{fig:self_propulsion} (a) Plots of $\mathcal L(t)/R_0$ [Eq.~(\ref{eq:spatem})] versus $(t-t_0)\nu/R_0^2$ for $A = 0$ (red curve), $A = 0.15$ (magenta curve), $A = 0.5$ (green curve), and and $A = 1$ (blue curve), with $t_0$ is a non-universal offset that depends on $A$. (b) Illustrative trajectories of the droplet's $CM$ for $A = 0.15$ (orange) and $A=1$ (blue-purple), with colorbars indicating the simulation time. (c) Log-log plots of the mean-square-displacement $\mathcal{M}(t)$ versus $t\nu/R_0^2$ (after the removal of initial transients) for droplet-$CM$ trajectories: $A=0.15$ (red), $A=0.5$ (green), $A=1$ (blue), $A=1.5$ (dark orange),  and $A=2$ (magenta);
initially these plots show ballistic regimes, but, at large times, we see $\mathcal{M}(t) \sim t^{\zeta}$, with $\zeta = 2$ (rectilinear motion for $A=0.15$), and superdiffusive regimes with $\zeta = 1.67 \pm 0.02 \simeq 5/3$ (for $A=0.5$) and $\zeta =1.28 \pm 0.05 \simeq 4/3$ (for $A=2$) via local-slope analysis (the inset shows plots of $\zeta$ versus $t$);  plots for different values of $A$ are displaced vertically for ease of visualization.} 
\end{figure*}
In Fig.~\ref{fig:phi_psi} we illustrate the evolution of the initial droplet [Eq.~(\ref{eq:init})] via pseudocolor plots of $\psi$ and the $\phi = 0$ contour (in magenta), at different representative times and $A = 0$ [row (a)], $A = 0.15$ [row (b)], and $A = 1$ [row (d)]; in row (c) we show, for $A = 0.15$, vector plots of the velocity field $\bm u$, with the $\phi = 0$ contour line, overlaid on a pseudocolor plot of the vorticity $\omega$ normalised by its maximal value. Case $A=0$ [row (a)]: there is no mean flow, i.e., $U_{CM}(t)=0$ for all $t$; however, as time increase (from left to right), the initially homogeneous mixture of active matter becomes unstable and undergoes phase separation via the formation of self-organized alternating rings of regions with positive and negative values of $\psi$ (cf. oil-water phase separation in a microfluidic droplet~\cite{moerman2018emulsion}); eventually, complete phase separation occurs, via successive ring collapses, and we obtain a $\psi>0$ region (red) surrounded by a $\psi < 0$ ring (blue) inside the $\phi=0$ contour.

\begin{figure*}
\hspace{-0.6cm}
{\includegraphics[width=18.5cm]{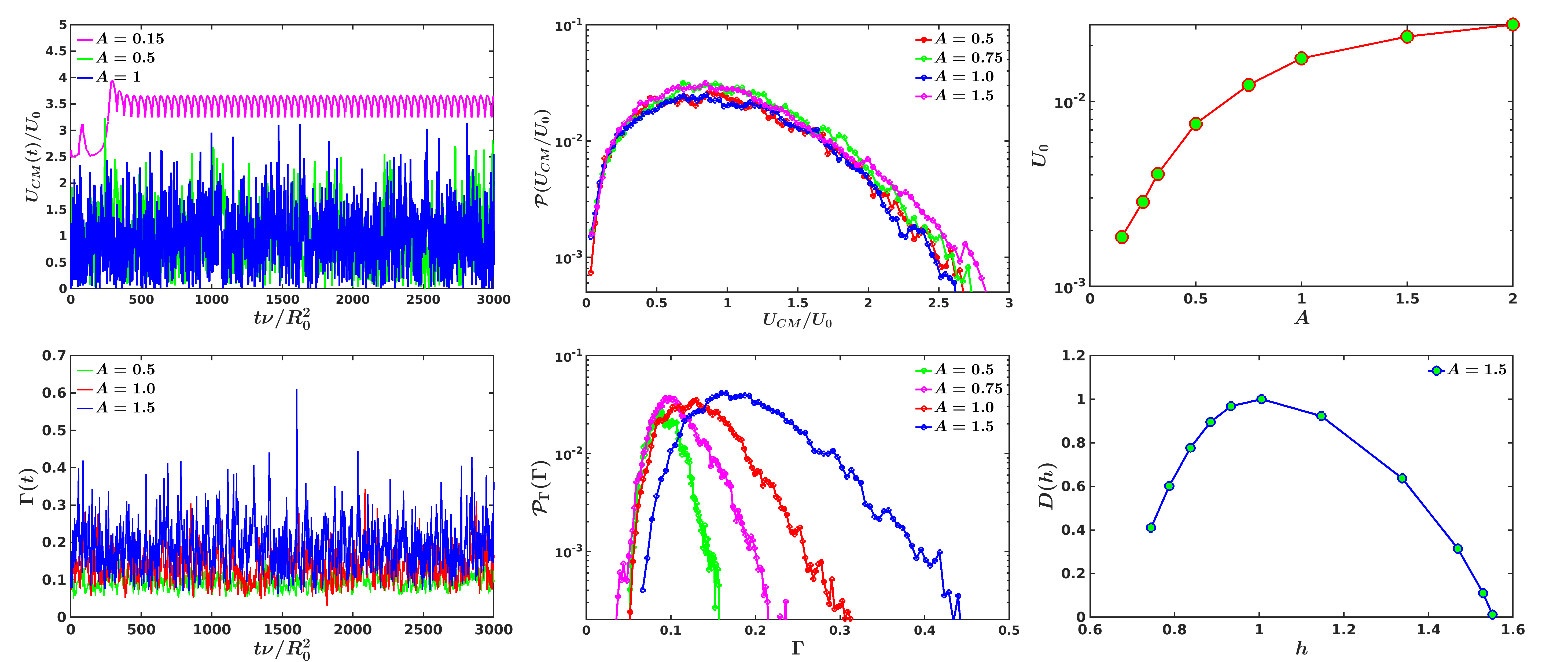}
\put(-525,205){\rm {\bf(a)}}
\put(-325,205){\rm {\bf(b)}}
\put(-120,205){\rm {\bf(c)}}
\put(-525,90){\rm {\bf(d)}}
\put(-325,90){\rm {\bf(e)}}
\put(-120,90){\rm {\bf(f)}}
}

\caption{\label{fig:vel_deform} (a) Plots versus the non-dimensionalized time $t\nu/R_0^2$ of the scaled droplet-$CM$ speed $U_{CM}/U_0$ [Eq.~(\ref{eq:spatem})] for $A = 0.15$ (magenta curve, which has been moved up to aid visualization), $A=0.5$ (green) and $A=1$ (blue). 
(b) Semilog plots of the PDF $\mathcal{P}(U_{CM}/U_0)$ for $A=0.5$ (red), $A=0.75$ (green), $A=1$ (blue), and $A=1.5$ (magenta). 
(c) Semilog plot of $U_0$ versus $A$. (d) Plots versus $t\nu/R_0^2$ of the normalised droplet perimeter $\Gamma(t)$ [Eq.~(\ref{eq:spatem})] for $A = 0.5$ (green), $A = 1$ (red), and $A = 1.5$ (blue). (e) Semilog plots of the PDF of  $\mathcal{P}_\Gamma(\Gamma)$ for $A = 0.5$ (green), $A = 0.75$ (magenta), $A = 1$ (red), and $A = 1.5$ (blue). 
 (f) Plots of the multifractal $D(h)$ versus the Hurst exponent $h$ (see text), obtained from $\Gamma(t)$, for $A = 1.5$.} 
\end{figure*}
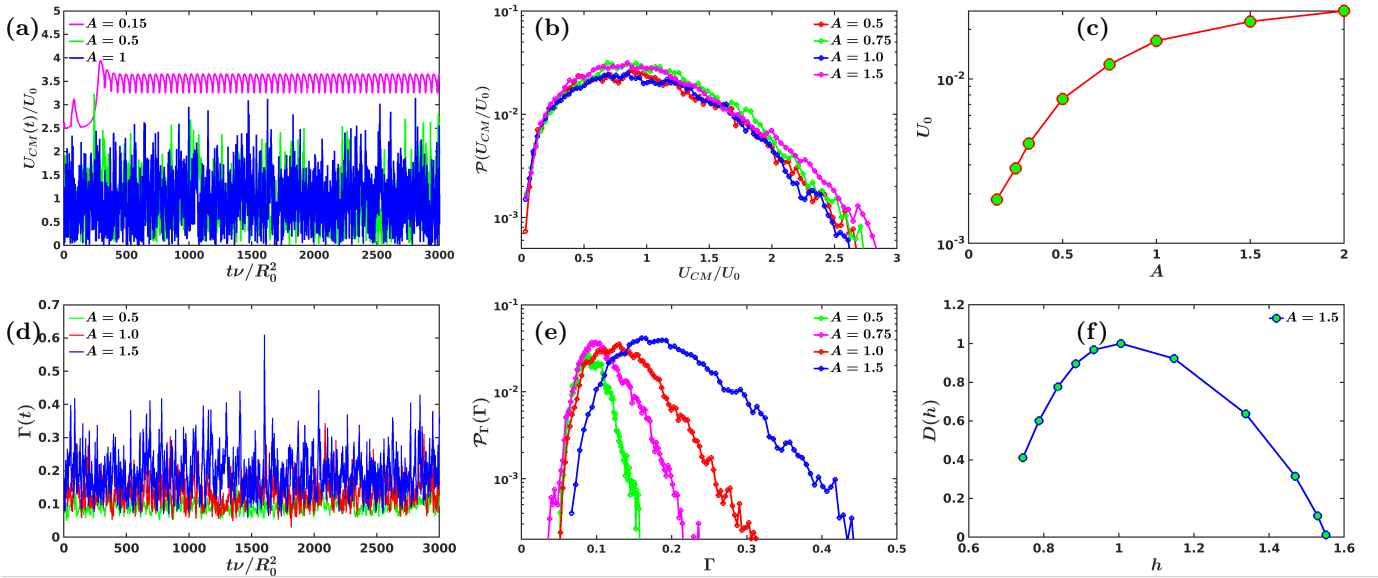

As we increase $A$, we find a remarkable transition to a \textit{self-propelling droplet}, whose motion we depict, for the illustrative value $A=0.15$, via pseudocolor plots in Figs.~\ref{fig:phi_psi}(b) and the video V2 in Appendix ~\ref{app:videos}. Initially, phase separation tries to set in, but is partially arrested; at this stage the flow field 
is dominated by a vortex quadrupole [second panels in Figs.~\ref{fig:phi_psi}(b) and (c)]; thereafter, an umbilicus, which forms at one end of the droplet, oscillates periodically in time as it shoots out a tiny blue bead, with $\psi < 0$ [third and fourth panels in Fig.~\ref{fig:phi_psi}(b)]; the associated flow patterns contain an oscillating vortex dipole [third and fourth panels in Fig.~\ref{fig:phi_psi} (c) and the video V3 in ~\ref{app:videos}] that propels the droplet along a straight line [the orange trajectory in Fig.~\ref{fig:self_propulsion} (b)]~\footnote{We have checked explicitly that the precise direction of droplet propulsion depends on the realization of the random distribution of $\psi_0 (\bm x)$ in the initial condition.}. These oscillations are mirrored in the periodic time dependence of $\mathcal L (t)/R_0$ [magenta curve in Fig.~\ref{fig:self_propulsion} (a)] and $U_{CM}(t)$ [magenta curve in Fig.~\ref{fig:vel_deform} (a)] and also in a limit cycle whose projection can be viewed in the $\mathcal L (t)\,-\,U_{CM}(t)$ plane [magenta curve in Fig. ~\ref{fig:supmat} (a) in Appendix ~\ref{app:non-dimensions}]. Thus, as we increase $A$, the initial transition from a static to a self-propelling droplet can be associated with the formation of a stable limit cycle. 

For sufficiently large $A\,(\gtrsim 0.5)$, statistically steady \textit{active-fluid turbulence} develops inside the droplet and leads to important modifications in its structure and propulsion: (a) We find a significant suppression of the phase separation of the active scalar [Figs.~\ref{fig:phi_psi}(d)], which is reminiscent of turbulence-induced coarsening arrest in a binary-fluid mixture~\cite{perlekar2017two}; (b) chaotic temporal fluctuations in $\mathcal L (t)/R_0$ [Fig.~\ref{fig:self_propulsion} (a)]; (c) convoluted trajectories of the $CM$ of the droplet [e.g., the blue-purple trajectory in Fig.~\ref{fig:self_propulsion} (b)], which are accompanied by chaotic temporal fluctuations in $U_{CM}(t)$ [Fig.~\ref{fig:vel_deform} (a)], and the projections of the phase-space trajectories in the $\mathcal L (t)\,-\,U_{CM}(t)$ plane [red and green curves in Fig. ~\ref{fig:supmat} (a) in Appendix ~\ref{app:non-dimensions}]; (d) multifractal fluctuations of $\Gamma(t)$ [Figs.~\ref{fig:vel_deform} (d)-(f)]; (e) the energy, $\phi$, and $\psi$ spectra that extend over a large range of the wave number $k$ [Figs. ~\ref{fig:spectra} (a)-(c) in Appendix ~\ref{app:non-dimensions}]. 

The transition from rectilinear to chaotic-droplet trajectories is apparent in the illustrative plots of droplet-$CM$ paths, for $A=0.15$ (orange) and $A=1$ (blue-purple) in Fig.~\ref{fig:self_propulsion} (b),
which we compute as in Ref.~\cite{bai2008calculating} (see Appendix ~\ref{app:droplet_cm}). From such paths we obtain the normalised mean-square-displacement $\mathcal{M}(t)/R^2_0$, which we present in log-log plots versus the non-dimensional time $t$ in Fig.~\ref{fig:self_propulsion} (c): rectilinear droplet motion leads to $\mathcal{M}(t) \sim t^2$ (red curve for $A=0.15$). As we increase $A$, we obtain crossovers to super-diffusive behaviors, which are consistent with $\mathcal{M}(t) \sim t^{5/3}$ (green curve for $A=0.5$) and $\mathcal{M}(t) \sim t^{4/3}$ (magenta curve for $A=2$), which suggest L\'evy walks for the droplet's $CM$ (cf. the motion of Lagrangian tracers in 2D turbulence and in a model for bacterial turbulence~\cite{elhmaidi1993elementary,mukherjee2021anomalous}).
The activity-induced transition from rectilinear-to-chaotic droplet motion is also mirrored in the time-dependence of $U_{CM}(t)/U_0$ that we depict in Fig.~\ref{fig:vel_deform}(a): the oscillatory behavior at $A=0.15$ (magenta curve) gives way to chaotic times series as we move from $A = 0.5$ (green)  to $A=1$ (blue). We characterize these chaotic fluctuations by computing the PDF $\mathcal{P}(U_{CM}/U_0)$, which we show in the semi-log plots of Fig.~\ref{fig:vel_deform}(b); these PDFs collapse onto each other, for different values of $A$, because we use the scaled speed $U_{CM}/U_0$; if we use the unscaled $U_{CM}$, then the skewness of this PDF increases with $A$ (Fig. ~\ref{fig:supmat} (c) in Appendix ~\ref{app:non-dimensions}). Furthermore, $U_0$ increases monotonically with $A$ [Fig.~\ref{fig:vel_deform} (c)] and shows signs of saturation at large $A$.

Not only does the active droplet display an increase in $U_0$ with $A$, but it also exhibits, as $A$ increases, an enhancement in fluctuations in its normalised perimeter $\Gamma(t)$ [Eq.~(\ref{eq:spatem})], which we plot versus $t\nu/R_0^2$ in Fig.~\ref{fig:vel_deform} (d). These fluctuations of $\Gamma$ lead to broad PDFs, $\mathcal{P}_\Gamma(\Gamma)$, which we present in Fig.~\ref{fig:vel_deform} (e), for $A = 0.5$ (green), $A = 0.75$ (magenta), $A = 1$ (red), and $A = 1.5$ (blue); the widths and skewnesses of these PDFs increase with $A$ (see Appendix ~\ref{app:non-dimensions}). From a multifractal analysis of the time series $\Gamma(t)$, we obtain the generalised spectrum of dimensions $D(h)$ as a function of the Hurst exponent $h$ by using the wavelet-leader method (see Refs.~\cite{jaffard2006wavelet,wendt2007multifractality} and Appendix ~\ref{app:multifractal}). In  Fig.~\ref{fig:vel_deform} (f) we present an illustrative plot of the multifractal spectrum $D(h)$ for $A=1.5$ (blue curve). Such multifractality has not been obtained heretofore for active droplets; it is akin to the recently discovered droplet-perimeter fluctuations in turbulent binary-fluid flows~\cite{pal2016binary}.  

As the activity induces turbulence in the $\psi$ field, the droplet's motion yields fluid turbulence, which we characterize by the energy, $\phi$, and $\psi$ spectra $E(k,t)$, $S_\phi(k,t)$, and $S_\psi(k,t)$ [Eq.~(\ref{eq:spatem})], that we plot in Figs. ~\ref{fig:spectra}(a)-(c) of Appendix ~\ref{app:non-dimensions} for $A = 0.5,\,1,$ and $A=1.5$. Even though the Reynolds numbers are small, these spectra span several decades in $k$, a clear signature of turbulence. We will present elsewhere~\cite{withKiran} a detailed study of the properties of a statistically homogeneous and isotropic form of this turbulence, which is reminiscent of bacterial or active-fluid turbulence~\cite{alert2022active, kiran2022irreversiblity, mukherjee2021anomalous} and elastic turbulence in polymer solutions~\cite{groisman2000elastic, gupta2017melting}.

\section{Conclusion}
We have developed a minimal model for assemblies of contractile swimmers, without alignment interactions, encapsulated in a droplet of a binary-fluid emulsion. Our hydrodynamic model, with the scalar order parameter $\phi$ and the active scalar $\psi$ coupled to each other and the velocity field $\bm u$, not only captures the droplet 
interface (via the $\phi = 0$ contour) and its fluctuations, but also leads to droplet self-propulsion, which is rectilinear at low $A (\simeq 0.15)$ and chaotic for large values of $A$, at which the $CM$ of the droplet shows super-diffusive motion and 
the droplet interface exhibits multifractal fluctuations. Our study is distinct from earlier theoretical studies of active droplets that consider cell-level models in nematically ordered, active polar, chemically driven, or phase-field systems~\cite{gao2017self,vcopar2019topology,tjhung2012spontaneous,de2014spontaneous,a2014active,ruske2021morphology,fadda2017lattice,singh2020self}. By contrast, the activity-induced droplet propulsion in our model arises from the interplay of $\phi$, $\bm u$, and a collection of contractile swimmers, which are described via the field $\psi$ and are enclosed inside the droplet;
this propulsion shows a hitherto unexplored crossover from
rectilinear to superdiffusive motion of the droplet $CM$. 
We look forward to the experimental verification of our results, especially in active droplets of contractile swimmers such as \textit{C. reinhardtii}~\cite{yeomans2014introduction,fragkopoulos2021self}, where it should be possible to control the activity by changing the oxygen concentration in low-light conditions.

\begin{acknowledgments}
 We thank J.K. Alageshan, K.V. Kiran, S.J. Kole, and S. Ramaswamy for discussions, the National Supercomputing Mission (NSM), and SERB (India) for support, and SERC (IISc) for computational resources.
\end{acknowledgments}
\appendix
\begin{widetext}
\section{Cahn-Hilliard formalism}
\label{app:chns_formalism}

It is convenient  to write the Cahn-Hilliard free-energy functional for a binary-fluid mixture in the following way:
\begin{eqnarray}
\mathcal F(\phi) = \int_{\Omega} \left[\frac{3}{16} \frac{\sigma}{\epsilon}(\phi^2-1)^2 + \frac{3}{4} \sigma \epsilon |\nabla \phi|^2\right]\,;
\end{eqnarray}
this depends on two important physical parameters: $\sigma$, the surface tension coefficient, and $\epsilon$, the interface width. The first term is a double-well potential with two minima at $\phi = \pm 1$. The equilibrium interfacial 
profile $\phi_0(x)$ can computed be computed by solving the following boundary value problem (if $\phi$
is assumed to vary along one spatial direction, say $x$):
\begin{eqnarray}
\begin{cases}
\mu(x) = -\frac{3}{2}\sigma \epsilon \frac{d^2\phi_0}{dx^2} + \frac{3}{4} \frac{\sigma}{\epsilon} (\phi_0^3 - \phi_0) = 0 \,;\\
\displaystyle{\lim_{x \to \pm \infty}} \phi_0 = \pm 1 \,.
\end{cases} 
\end{eqnarray}
Here, $\mu(x) \equiv \frac{\delta \mathcal F}{\delta \phi}$ is the chemical potential. The solution is
\begin{eqnarray}
\phi_0(x) = \tanh{\left(\frac{x - x_0}{\epsilon} \right)}\,,
\end{eqnarray}
where $x_0$ is the midpoint of the interface. The interfacial free-energy is
\begin{eqnarray}
\displaystyle \int_{-\infty}^{\infty} \mathcal F(\phi_0) dx = \sigma \,.
\end{eqnarray}
The advantage of writing the free-energy functional in the above form is that $\sigma$ and $\epsilon$ can be varied independently of each other. In our direct numerical simulations, we tune the value of $\epsilon$, depending upon the computational mesh size, without changing the surface tension $\sigma$.

\textit{The stress terms:} In the passive CHNS or model-H equations, the stress in the Navier-Stokes equations is
\begin{eqnarray}
\mathcal {\mathfrak{S}}^{\phi} &=& -\phi \nabla \mu = \mu \nabla \phi - \nabla (\phi \mu) \nonumber \\
&=& -\frac{3}{2} \sigma \epsilon \nabla^2 \phi \nabla \phi - \nabla \left(\phi \mu - \frac{3}{4} \frac{\sigma}{\epsilon} (\phi^4/4 - \phi^2/2)\right) \,.
\end{eqnarray}
The second term on the right-hand side of the above equation vanishes when we take the curl to obtain the vorticity equation, so we use the stress $\mathfrak{S}^{\phi} = -(3/2) \sigma \epsilon \nabla^2 \phi \nabla \phi$. 
There is an alternative way of writing this stress, in terms of a stress tensor, as follows:
\begin{eqnarray}
\mathfrak{\bm S^\phi} &=& \nabla \cdot \Sigma^\phi = -(3/2) \sigma \epsilon \nabla^2 \phi \nabla \phi \,;\nonumber \\
\Sigma_{ij}^\phi &=&  -(3/2) \sigma \epsilon \left[(\partial_i \phi)(\partial_j \phi) - |\nabla \phi|^2 \delta_{ij}/2\right] \,.
\end{eqnarray}
Similarly, for the stress terms in our active CHNS model, in the main paper, we use:
\begin{eqnarray}
\mathfrak{S}^{\phi} &=& -(3/2)\sigma_1 \epsilon_1 \nabla^2 \phi \nabla \phi \,;\label{eq:S_phi}\\
 \mathfrak{S}^{\psi} &=&    -(3/2) \tilde\sigma_2 \epsilon_2 \nabla^2 \psi \nabla \psi\,. \label{eq:S_psi}
\end{eqnarray}

\section{Non-dimensional forms}
\label{app:non-dimensions}

We write the non-dimensionalized forms of Eqs.~(1)-(7), in the main paper, by using the following transformations:
\begin{eqnarray}
\bm x^* = \bm x / R_0\,; \ \bm u^* = \bm u / U_0\,; \ t^* = t U_0/R_0\,; \ \omega^* = \omega U_0/R_0\,. 
\end{eqnarray}
The non-dimensionalized equations are (we drop the superscript * to simplify the notations):
\begin{eqnarray}
\partial_t \omega + (\bm u \cdot \nabla) \omega &=& \frac{1}{\text{Re}} \nabla^2 \omega -\alpha^{\prime} \omega - \frac{3}{2} \frac{\text{Cn}_1}{\text{We}_1} [\nabla \times (\nabla^2 \phi \nabla \phi)] - \frac{3}{2} \frac{\text{A} \text{Cn}_2}{\text{We}_2} [\nabla \times (\nabla^2 \psi \nabla \psi)]\,;\\
\partial_t \phi + (\bm u \cdot \nabla) \phi &=& \frac{3}{2\text{Pe}_1} \nabla^2\left[-\text{Cn}_1^2  \nabla^2\phi + \frac{1}{2} (\phi^3 - \phi) - \frac{2}{3}\beta_1^{\prime} \psi\right]\,;\\ 
\partial_t \psi + (\bm u \cdot \nabla) \psi &=& \frac{3}{2\text{Pe}_2} \nabla^2\left[-\text{Cn}_2^2  \nabla^2\psi + \frac{1}{2} (\psi^3 - \psi) - \frac{2}{3}\beta_2^{\prime} \phi\right]\,.
\end{eqnarray}

The important \textit{non-dimensional numbers} are:
\begin{eqnarray}
\text{the Reynolds number:}\ \text{Re} &=& R_0U_0/\nu\,; \nonumber \\
\text{the non-dimensionalized friction:} \ \alpha^{\prime} &=& \alpha R_0 / U_0\,; \nonumber \\
\text{Cahn numbers:}\ \text{Cn}_1 &=& \epsilon_1/R_0\,; \, \text{Cn}_2 = \epsilon_2 / R_0\,; \nonumber \\ 
\text{Weber numbers:} \ \text{We}_1 &=& R_0 U_0^2 / \sigma_1\,; \, \text{We}_2 = R_0 U_0^2/\sigma_2\,; \nonumber \\
\text{Activity parameter:} \ \text{A} &=& |\tilde \sigma_2|/\sigma_2\,; \\
\text{Order-parameter couplings:}\ \beta_1^{\prime} &=& \beta \epsilon_1/\sigma_1\,; \, \beta_2^{\prime} = \beta \epsilon_2 / \sigma_2\,; \nonumber \\
\text{Peclet numbers.}\ \text{Pe}_1 &=& \frac{R_0 U_0 \epsilon_1}{M_1 \sigma_1}\,; \text{Pe}_2 = \frac{R_0 U_0\epsilon_2}{M_2 \sigma_2}\,. \nonumber  
\end{eqnarray}

\begin{table}[H]
		\centering
		\begin{tabular}{| c | c | c | c | c | c |}
			\hline
			Run & $A$ &  Re & $\alpha^{\prime}$ & $\text{We}_1 = \text{We}_2$ & $\text{Pe}_1 = \text{Pe}_2$\\
			\hline
			R1& 0 & 0 & $\infty$ & 0 & 0\\
			\hline
			R2& 0.15 & 0.002 & 85 &$\mathcal O(10^{-6})$ & 0.1\\
			\hline
			R3& 0.5 & 0.008 & 20 & $\mathcal O(10^{-5})$ & 0.4 \\
			\hline
			R4& 0.75 & 0.013 & 13 & $\mathcal O(10^{-4})$& 0.7\\
			\hline
			R5& 1.0 & 0.018 & 9 & $\mathcal O(10^{-4})$ & 1.0\\
			\hline
			R6& 1.5 & 0.02 & 7 & $\mathcal O(10^{-4})$ & 1.3\\
			\hline
			R7& 2 & 0.03 & 6 & $\mathcal O(10^{-3})$ & 1.5 \\
			\hline
		\end{tabular}
		\caption{\label{tab:param} The values of various non-dimensional parameters in our DNS Runs R1-R7.
		The following parameters are fixed in all these runs:
        $N = 512, \; \text{grid size}, dx = 2\pi/N, \; R_0 = \pi/2, \; \text{Cn}_1 = 3dx/R_0 = \text{Cn}_2, \; \beta_1^{\prime} =  0.075 = \beta_2^{\prime}, \; M_1 = 10^{-3} = M_2, \; \sigma_1 = 1 = \sigma_2.$}
		\label{tab:parameters}
\end{table}
\begin{figure*}
{
\includegraphics[width=18cm]{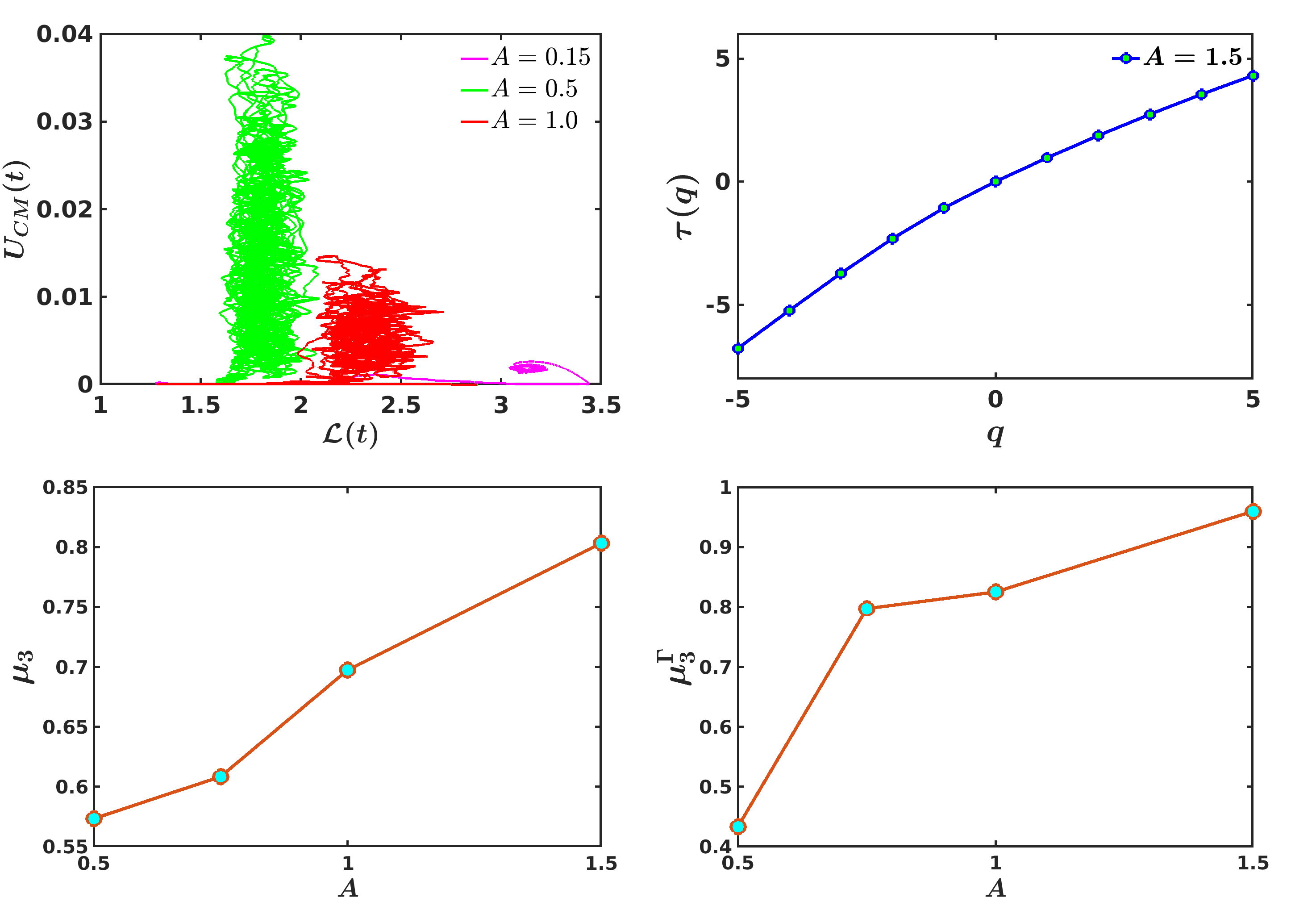}
\put(-460,310){\rm {\bf(a)}}
\put(-220,310){\rm {\bf(b)}}
\put(-460,140){\rm {\bf(c)}}
\put(-220,140){\rm {\bf(d)}}
}
\caption{\label{fig:supmat} (a) For $A = 0.15$, the system shows oscillatory behavior; this is confirmed by phase-space 
trajectories settling onto a limit cycle (the projection in the $U_{CM} - \mathcal{L}$ plane is shown via the illustrative magenta trajectory); as the system becomes turbulent, these trajectories become chaotic (see, e.g., the representative trajectories for $A = 0.5$ (green) and $A = 1$ (red). (b) Plot of the generalized exponent $\tau(q)$ as a function of the order $q$ for the
representative value $A=1.5$; the deviation from the linearity suggests the multifractality of $\Gamma(t)$. Plots versus $A$ of (c) the skewness $\mu_3$ of the PDFs of $U_{CM}(t)$ and (d) the skewness $\mu^{\Gamma}_3$ of the PDFs of $\Gamma(t)$.}
\end{figure*}
\begin{figure*}
{\includegraphics[width=18.5cm]{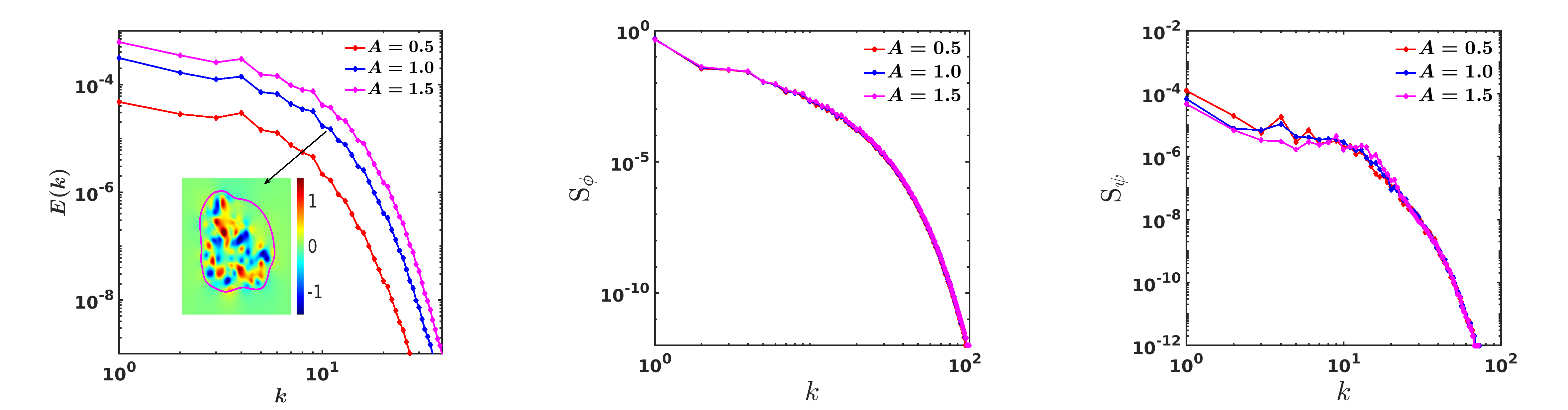}
\put(-525,130){\rm {\bf(a)}}
\put(-350,130){\rm {\bf(b)}}
\put(-170,130){\rm {\bf(c)}}
}
\caption{\label{fig:spectra} Log-log plots versus the wave number $k$ of the spectra (a) $E(k)$, (b) $\text{S}_\phi(k, t)$,  and (c) $\text{S}_\psi(k,t)$ [Eq.10 in the main text], averaged over time in the statistically steady state, for $A = 0.5$ (red curves), $A = 1$ (blue curves), and $A = 1.5$ (magenta curves); the inset in (a) shows a representative pseudocolor plot of the vorticity 
field for $A = 1$.}
\end{figure*}
\section{Multifractal anlysis}
\label{app:multifractal}

We use the wavelet-leader technique~\cite{jaffard2006wavelet,wendt2007multifractality} to obtain the multifractal spectrum 
of the Hurst exponents for the time series $\Gamma(t)$; we  employ \textit{MATLAB}~\cite{MATLAB} to perform this 
multifractal analysis. The wavelet leader $T_{\psi}[f](t, a)$ is obtained from the convolution operation for a time series 
$f(t)$  as follows:
\begin{eqnarray}
T_{\chi}[f](t, a) = \frac{1}{a} \int_{-\infty}^{+\infty} \chi\left(\frac{t - b}{a}\right) f(t) dt\,.
\end{eqnarray}
Here, $\chi(t)$ is the form for a single wavelet, with $a$ and $b$ the scale and translation parameters, respectively. The structure function of order $q$, based on this wavelet leader, is 
\begin{eqnarray}
S_{T}(q, a) = \frac{1}{N_w}\displaystyle \sum_{i=1}^{N_w} |T_{\chi}[f](t, a)|^q \sim a^{\tau(q)}\,,
\end{eqnarray}
with $N_w = n/a$ the number of wavelets for a particular scale $a$ and $n$ the length of the entire time series. The Legendre transform of the generalized exponents $\tau(q)$ gives an upper bound for the generalized dimensions $D(h)$:
\begin{eqnarray}
D(h) \le \displaystyle \min_{q \ne 0} [1 + qh(q) - \tau(q)]\,,
\end{eqnarray}
where $h(q) \equiv d\tau(q)/dq$ is the H\"{o}lder exponent of order $q$. For a monofractal time series, $\tau(q)$ is a linear function of $q$, so $h$ is independent of $q$ and $D(h)$ is nonzero for only this value of $h$. By contrast, for a multifractal time series, $\tau(q)$ is a nonlinear function of $q$ [Fig.~\ref{fig:supmat}(b)]  and $D(h)$ is a nontrivial function of $h$
(see the main paper).

\section{Calculation of the droplet's center-of-mass $(CM)$}
\label{app:droplet_cm}

The Fourier transform of the field $\phi(x, y)$ is 
\begin{equation}
\hat\phi(k_x, k_y) = \displaystyle \sum_{(x, y)} \phi(x, y) \exp^{\imath (k_x x + k_y y)}\,.    
\end{equation}
The two components of the position of the droplet's $CM$ can now be calculated as follows~\cite{bai2008calculating}:
\begin{eqnarray}
X_{CM} &=& \arctan\left(\Im (\hat \phi(1, 0)), \Re (\hat \phi(1, 0))\right)\,; \nonumber \\
Y_{CM} &=& \arctan\left(\Im (\hat \phi(0, 1)), \Re (\hat \phi(0, 1))\right)\,. 
\end{eqnarray}

\section{Videos}
\label{app:videos}
The following videos are available on request; please send an email to nadia@iisc.ac.in and rahul@iisc.ac.in.
\begin{enumerate}
\item Video V1: Video showing the spatiotemporal
evolution of the pseudocolor plots in  Fig. 1 of the main text (row (a) for $A = 0$) and illustrating that, when $A=0$, the domain growth of the $\psi$-field is driven purely by diffusion.
\item Video V2: Video showing the spatiotemporal
evolution of the pseudocolor plots in  Fig. 1 of the main text (row (b) for $A = 0.15$).
\item Video V3: Video showing the spatiotemporal evolution of vector plots of the velocity field $\bm u$, with the $\phi = 0$ contour line (magenta), overlaid on a pseudocolor plot of the vorticity $\omega$ normalised by its maximal value
(cf. Fig. 1 of the main text (row (c) for $A = 0.15$)); the lengths of the velocity vectors are proportional to their magnitudes.
\item Video V4: Video showing the spatiotemporal
evolution of the pseudocolor plots in  Fig. 1 of the main text (row (d) for $A = 1$).
\end{enumerate}

\end{widetext}
\nocite{*}
\bibliographystyle{apsrev4-2}
%


\end{document}